\documentclass[preprint,showpacs,preprintnumbers,amsmath,amssymb,aip]{revtex4-1}

\usepackage{graphicx}
\usepackage{dcolumn}
\usepackage{bm}
\usepackage{psfrag}

\begin{document}

\preprint{APS/123-QED}

\title{Scaling laws and bounds for the turbulent G.O. Roberts dynamo}

\author{A. Tilgner}

\affiliation{Institute of Geophysics, University of G\"ottingen,
Friedrich-Hund-Platz 1, 37077 G\"ottingen, Germany }

\date{\today}

\begin{abstract}
Numerical simulations of the G.O. Roberts dynamo are presented. Dynamos both
with and without a significant mean field are obtained. Exact bounds are derived
for the total energy which conform with the Kolmogorov phenomenology of
turbulence. Best fits to numerical data show the same functional dependences 
as the inequalities obtained from optimum theory.
\end{abstract}

\maketitle

\section{Introduction}

Purely hydrodynamic or magnetohydrodynamic (MHD) flows driven by a volume force
in three dimensional periodic boxes have a long history in the study of
fundamental properties of turbulence and magnetic field generation. Under the
assumption that there are universal features in turbulence independent of the
shape of distant boundaries, spatially periodic models should allow us to study
those features conveniently. Body forces are frequently used to model the effect
of actual boundaries, such as moving propellers, which are too cumbersome to 
simulate in full. While this problem is mostly rooted in engineering, it has
also become of interest in MHD research because of laboratory dynamo experiments
in which liquid metal is forced by a moving boundary more easily represented by
a volume force in numerical simulations intended to design or reproduce the
experiments (as for example in \cite{Baylis07,Reuter09}). Finally, some
mechanisms to drive flows in geo- or astrophysics involve a volume forcing if
observed in the appropriate frame of reference, such as precession or tides
\cite{Tilgne15}.

Even the numerically convenient setting of a flow in a periodic box overstrains
computational possibilities if Reynolds numbers get too large. Exact analytical
solutions for turbulent flows are unknown, but techniques are known to determine
at least upper bounds for certain quantities. This approach is frequently
called optimum theory. These techniques have been applied
to several idealized flows, notably Couette flows and convection in plane
layers, and also to MHD problems (\cite{Soward80,Kerswe96,Albous09} and
references therein). These bounds were obtained by using either a background
field technique going back to Hopf \cite{Hopf41} and Doering and Constantin \cite{Doerin92} 
, or by using an optimization
procedure introduced by Howard \cite{Howard72} and later on championed by Busse \cite{Busse78}, mostly
in the context of convection. Both of these methods rely on the presence of
boundaries where boundary conditions restrict admissible fields, and are not
useful in a periodic domain. The results presented here for dynamos
draw on the work for non-magnetic flows 
in periodic boxes by Doering and Foias \cite{Doerin02} (see also
\cite{Childr01,Rollin11}) who formulate bounds in as general terms as possible. The
present paper proceeds by way of example and studies a periodic flow
introduced by G.O. Roberts \cite{Robert72}, but the procedure is applicable to general
forcing. The G.O. Roberts flow has served as paradigm for numerous problems
\cite{Tilgne07d,Tilgne08,Tilgne08b,Ponty11,Tanriv11} and resembles qualitatively
the helical flows generated by convection in rotating systems.

The model which will be considered here is a three dimensional periodic box
driven by a force density such that the flow assumed in the G.O. Roberts dynamo
is a solution of the Navier-Stokes equation with that forcing. In general, this
flow is unstable. If magnetic field growth sets in, the most important question
is at which amplitude the field will saturate. This issue was already tackled
with the tools of weakly nonlinear analysis \cite{Petrel01,Tilgne02b}. Far from
onset, dimensional or heuristic arguments are necessary
\cite{Brande01,Petrel07}. Heuristics are always uncertain, and dimensional
arguments face the problem that more than one length scale (different
periodicity lengths in different directions) or more than one velocity scale
(the kinetic velocity and the Alfv\`en velocity) may be available to form
expressions with certain prescribed units. Optimum or bounding theory on the
other hand is rigorous, but has a mixed record regarding its performance in
approximating or tightly bounding the quantity of interest. A bound is of course
the more interesting the closer it is to the value it bounds. The present paper
explores the possibilities of this tool when applied to the G.O. Roberts dynamo.

\section{The model}

Consider a fluid with density $\rho$, conductivity $\sigma$, kinematic viscosity
$\nu$, magnetic permeability $\mu_0$ and magnetic diffusivity $\lambda=\mu_0
\sigma$ moving at a velocity $\bm v(\bm r,t)$ as a function of position $\bm r$
and time $t$. The magnetic field is $\bm B(\bm r,t)$. The flow is driven by the
time independent force density $\bm f(\bm r)$. Periodic boundary conditions are
imposed with a periodicity length $h$ along the $z-$axis and $a$ along the $x-$
and $y-$axes. It will later be of interest to apply additional restrictions and
to allow mean fields (obtained by averaging over the periodicity volume or over
planes) only for certain components. Dimensions are removed from the equations
of evolution by using $a$ and $a^2/\lambda$ as units of length and time, and by
introducing the rescaled variables
$\bm v a/\lambda$, $p a^2/(\nu \lambda \rho)$ ($p$ is the pressure),
$\bm B a/\sqrt{\rho \mu_0 \lambda \nu}$, and $\bm f a^3/(\rho \lambda \nu)$.
Using the same symbols for the rescaled and for the original variables, the
equations of evolution read:

\begin{eqnarray}
\frac{1}{\mathrm{Pm}} \left( \partial_t\bm v + \bm v \cdot \nabla \bm v \right)
&=& -\nabla p + \nabla^2 \bm v + (\nabla \times \bm B) \times \bm B + \bm f
\label{NS}\\
\partial_t\bm B + \nabla \times (\bm B \times \bm v) &=&  \nabla^2 \bm B 
\label{eq_induc}\\
\nabla \cdot \bm v = \nabla \cdot \bm B &=& 0 
\label{eq_div0}
\end{eqnarray}

with $\mathrm{Pm}=\nu/\lambda$. The periodicity enforces
$\bm v(\bm r,t)= \bm v(\bm r+ \bm {\hat x},t) = \bm v(\bm r+ \bm {\hat y},t) =
\bm v(\bm r+ l_z \bm{\hat z},t)$ with $l_z=h/a$ and likewise for $\bm B$. Hats
denote unit vectors.

With $V$ the volume of the periodicity cell, define time averaged densities of
magnetic energy $E_B$ and dissipation $\epsilon_B$ as
\begin{equation}
E_B= \langle \frac{1}{V} \int \frac{1}{2} \bm B^2 dV \rangle
~~~,~~~
\epsilon_B = \langle \frac{1}{V} \int (\nabla \times \bm B)^2 dV \rangle
\end{equation}
and time averaged densities of kinetic energy and dissipation as
\begin{equation}
E_{\mathrm{kin}}= \langle \frac{1}{V} \int \frac{1}{2} \bm v^2 dV \rangle
~~~,~~~
\epsilon_v = \langle \frac{1}{V} \int (\partial_j v_i) (\partial_j v_i) dV
\rangle= 
\langle \frac{1}{V} \int (\nabla \times \bm v)^2 dV \rangle,
\end{equation}
where $\langle ... \rangle$ denotes time average and with the standard index
notation in which summation over repeated indices is implied.
It will be convenient to use the force amplitude $F$ defined through
$\bm f(\bm r) = F ~ \bm \Phi(\bm r)$, with $F \ge 0$. The shape function
$\bm \Phi$ is normalized to
$\frac{1}{V} \int |\bm \Phi(\bm r)|^2 dV = 1$
and obeys $\nabla \cdot \bm \Phi = 0$ (a contribution with non zero divergence
to $\bm f$ can always be balanced by the pressure gradient).

A characteristic velocity can be defined from the kinetic energy. Because
velocities are given in multiples of $\lambda/a$ in the units chosen here, this
characteristic velocity actually is the magnetic Reynolds number $\mathrm{Rm}$
defined as
\begin{equation}
\mathrm{Rm}=\sqrt{2 E_{\mathrm{kin}}}.
\label{eq_Rm_def}
\end{equation}
The hydrodynamic Reynolds number $\mathrm{Re}$ is given by
\begin{equation}
\mathrm{Re}=\mathrm{Rm}/\mathrm{Pm}.
\end{equation}
The choice of units also leads to a factor $\mathrm{Pm}$ in the expression for
the total energy $E_{\mathrm{tot}}$:
\begin{equation}
E_{\mathrm{tot}} = \frac{1}{\mathrm{Pm}} E_{\mathrm{kin}} + E_B
\label{eq_Etot_def}
\end{equation}

One obtains an energy balance by dotting eq. (\ref{NS}) with $\bm v$, eq.
(\ref{eq_induc}) with $\bm B$, integrating over $V$, adding the two equations
and using the fact that time derivatives of time averages are zero:
\begin{equation}
\epsilon_v + \epsilon_B = F \langle \frac{1}{V} \int \bm \Phi \cdot \bm v ~ dV
\rangle
\label{eq_budget}
\end{equation}

Equations (\ref{NS}-\ref{eq_div0}) were solved numerically for a volume force of
the form
\begin{eqnarray}
\bm \Phi=
\left( \begin{array}{ll}
\sin (2 \pi x) \cos (2 \pi y)\\
-\cos (2 \pi x) \sin (2 \pi y)\\
\sqrt{2} \sin (2 \pi x) \sin (2 \pi y)\\
\end{array} \right)
\label{eq_flowI}
\end{eqnarray}
in the domain $0\le x,y \le 1$, $0\le z \le l_z$. Velocity and magnetic fields were enforced to
be periodic in all three directions. Zero wavenumber modes were allowed for $\bm
v$ in the $z-$direction (to conform with the structure of $\bm \Phi$) and for
$\bm B$ in the $x-$ and $y-$directions (to allow a mean field in the
$x,y-$plane).

The results of six suites of simulations will be presented below, three for
$l_z=1$ and three for $l_z=2$. For both $l_z$, three series of runs simulated
$\mathrm{Pm}=3$, 1 and 0.3 for different $F$. The numerical scheme was the same
finite difference method implemented on GPUs as used in \cite{Tanriv11}.
Spatial resolutions reached up to $512^3$.

\section{Bounds}

All the bounds derived below follow from two theorems. The first is the Cauchy-Schwarz
inequality stating that
\begin{equation}
|\int g h ~ dV|^2 \le \int g^2 dV \cdot \int h^2 dV
\end{equation}
for any two square integrable functions $g$ and $h$, and the second is
Poincar\'e's inequality in the form
\begin{equation}
\int g^2 dV \le l^2 \int |\nabla g|^2 dV
\end{equation}
where $l^2$ is the inverse of the smallest eigenvalue of $-\nabla^2$ whose
eigenfunction is compatible with the boundary conditions and restrictions
imposed on $g$ \cite{Doerin02}. We will apply Poincar\'e's inequality to components of both
velocity and magnetic fields and use inverse eigenvalues $l_v^2$ and $l_B^2$.
Periodicity alone allows spatially constant fields, leading to infinite $l_v$
and $l_B$. However, we will want to exclude uniform translation and a uniform
magnetic field (which does not decay through ohmic diffusion), and $l_v$ and
$l_B$ are determined by the smallest wavevector of the admitted fields. It may
be interesting to make different choices for $\bm v$ and $\bm B$ regarding the
admissible fields with the largest length scale, as was done at the end of the
previous section, so that we keep track of two different lengths, $l_v$ and
$l_B$.

It is now trivial to obtain bounds in which $F$ plays the role of the control
parameter. Poincar\'e's inequality applied to the definitions of energies and
dissipation rates yields
\begin{equation}
\epsilon_v \ge \frac{1}{l_v^2} 2 E_{\mathrm{kin}}
~~~,~~~
\epsilon_B \ge \frac{1}{l_B^2} 2 E_B.
\label{eq_def_epsilon}
\end{equation}
Combined with the Cauchy-Schwarz inequality applied to the eq. (\ref{eq_budget}), this leads to
\begin{equation}
\frac{\mathrm{Rm}^2}{l_v^2} + \frac{2 E_B}{l_B^2} \le \epsilon_v + \epsilon_B
\le F ~ \mathrm{Rm}.
\label{eq_14}
\end{equation}
Since all variables in this inequality are positive, it immediately follows that
\begin{equation}
\mathrm{Rm} \le F ~ l_v^2
\end{equation}
and from searching the extremum of $E_B$ over $\mathrm{Rm}$ in eq. (\ref{eq_14}), that
\begin{equation}
E_B \le \frac{1}{8} (l_B l_v F)^2
\end{equation}
so that the total energy is bounded from above by
\begin{equation}
E_{\mathrm{tot}} = \frac{1}{\mathrm{Pm}} E_{\mathrm{kin}} + E_B \le
(\frac{1}{\mathrm{Pm}} \frac{1}{2} l_v^2 + \frac{1}{8} l_B^2) l_v^2 F^2.
\label{eq_F2_Etot}
\end{equation}

This upper bound is of very limited practical interest. The dependence of
$E_{\mathrm{tot}}$ in $F^2$ suggests that this bound needs to be so large
in order to accommodate laminar
solutions which will become unstable in an actual time integration at
sufficiently large $F$, so that the $E_{\mathrm{tot}}$ obtained in numerical
simulations of turbulent flows will be far below this bound. Optimum theory as
we currently know it is not able to discriminate between stable and unstable
solutions, which is one of its major weaknesses.

For now, we will seek bounds independent of $F$ which relate energies and
dissipation rates. One obtains from dotting eq. (\ref{NS}) with $\bm f$ and
averaging:
\begin{equation}
\frac{1}{\mathrm{Pm}} F \langle \frac{1}{V} \int \bm \Phi \cdot \{(\bm v \cdot
\nabla) \bm v\} dV \rangle
=
F \langle \frac{1}{V} \int \bm \Phi \cdot \nabla^2 \bm v dV \rangle +
F \langle \frac{1}{V} \int \bm \Phi \cdot \{ (\nabla \times \bm B) \times \bm B
\} dV \rangle + F^2
\label{eq_Ftest}
\end{equation}
Two of the integrals appearing in this equation can be manipulated into more
convenient forms:
\begin{equation}
\int \bm \Phi \cdot \{(\bm v \cdot \nabla) \bm v \} dV = 
-\int v_i v_j \frac{1}{2} (\partial_i \Phi_j +\partial_j \Phi_i) dV
\end{equation}
and
\begin{equation}
\int \bm \Phi \cdot \{ (\nabla \times \bm B) \times \bm B \} dV =
-\int B_i B_j \frac{1}{2} (\partial_i \Phi_j +\partial_j \Phi_i) dV
\end{equation}
The tensor $\frac{1}{2} (\partial_i \Phi_j +\partial_j \Phi_i)$ is real and
symmetric and therefore has real eigenvalues. Denote by $e$ the largest absolute
value of these eigenvalues found anywhere in the volume. One then has
\begin{equation}
|\langle \frac{1}{V} \int \bm \Phi \cdot \{(\bm v \cdot \nabla) \bm v \} dV
\rangle|
\le 2 e E_{\mathrm{kin}}
\end{equation}
and
\begin{equation}
|\langle \frac{1}{V} \int \bm \Phi \cdot \{ (\nabla \times \bm B) \times \bm B
\} dV \rangle|
\le 2 e E_B.
\end{equation}
These bounds could be sharpened. In order to improve the first inequality, one
could determine the divergence free, periodic field $\bm v$ which maximizes
$\int \bm \Phi \cdot \{(\bm v \cdot \nabla) \bm v \} dV$ and obtain a better
estimate from there. This procedure requires the solution of a multidimensional
linear eigenvalue problem, which is a large numerical effort which does not
seem appropriate for this paper. It should also be pointed out that in the
context of the G.O. Roberts flow, the first of the above integrals will be
grossly overestimated, anyway. It is known from numerical simulations that the
velocity field resembles the laminar flow even if the forcing is chosen large
enough to create turbulence \cite{Ponty11}. If $\bm \Phi$ is the force field of
the Roberts flow of eq. (\ref{eq_flowI}) and $\bm v$ the laminar response, then
the integral $\int \bm \Phi \cdot \{(\bm v \cdot \nabla) \bm v \} dV$ is
strictly zero.

The remaining integral in eq. (\ref{eq_Ftest}) can be estimated using the
Cauchy-Schwarz inequality:
\begin{eqnarray}
\langle \frac{1}{V} \int \bm \Phi \cdot \nabla^2 \bm v dV \rangle & = &
\langle \frac{1}{V} \int \bm v \cdot \nabla^2 \bm \Phi dV \rangle =
\frac{1}{V} \int \langle \bm v \rangle \cdot \nabla^2 \bm \Phi dV 
\le  \sqrt{\frac{1}{V} \int \langle \bm v \rangle^2 dV} \sqrt{\frac{1}{V} \int
|\nabla^2 \bm \Phi|^2 dV} 
\nonumber
\\
& \le &
\sqrt{\frac{1}{V} \langle \int \bm v^2 dV\rangle} ~ \| \nabla^2 \bm \Phi \| =
\mathrm{Rm} ~ \| \nabla^2 \bm \Phi \| 
\end{eqnarray}
with the shorthand notation 
$\| \nabla^2 \bm \Phi \| = \sqrt{\frac{1}{V} \int |\nabla^2 \bm \Phi|^2 dV}$.
Inserting these inequalities into eq. (\ref{eq_Ftest}) leads to
\begin{equation}
F \le 2 e E_{\mathrm{tot}} + \| \nabla^2 \bm \Phi \| ~ \mathrm{Rm}.
\label{eq_F_Etot}
\end{equation}
This inequality is finally introduced into
$\epsilon_v + \epsilon_B \le F ~ \mathrm{Rm}$
(see eq. (\ref{eq_14})) to yield
\begin{equation}
\epsilon_v + \epsilon_B \le 2 e E_{\mathrm{tot}} \mathrm{Rm} + \| \nabla^2 \bm
\Phi \| ~ \mathrm{Rm}^2
\label{eq_epsilon_Etot}
\end{equation}
It can be seen from the definitions of $\mathrm{Rm}$ and $E_{\mathrm{tot}}$,
eqs. (\ref{eq_Rm_def}) and (\ref{eq_Etot_def}), that $E_{\mathrm{tot}}$
varies as a function of $\mathrm{Rm}$ as $\mathrm{Rm}^2$ for
$E_{B}=0$ and possibly faster if $E_{B} \neq 0$.
The first term on the right hand sides of eqs. (\ref{eq_F_Etot})
and (\ref{eq_epsilon_Etot}) thus exceeds the second term by at least
a factor $\mathrm{Rm}$.
If we now restrict attention to the limit $\mathrm{Rm} \rightarrow \infty$ and
retain only the dominating term in $\mathrm{Rm}$ on the right hand sides
in eqs. (\ref{eq_F_Etot})
and (\ref{eq_epsilon_Etot}), we find that asymptotically,
\begin{equation}
E_{\mathrm{tot}} \ge \frac{1}{2e} F
\end{equation}
and
\begin{equation}
\epsilon_v + \epsilon_B \le 2 e E_{\mathrm{tot}} \mathrm{Rm}.
\label{eq_Kolmogo_Etot}
\end{equation}
The first inequality is a lower bound for the total energy in terms of $F$ which
complements the upper bound derived above and expresses the scaling expected for
turbulent flow.

The second inequality (\ref{eq_Kolmogo_Etot}), if used as an equality,
is familiar from the Kolmogorov
picture of turbulence, in that it states that a dissipation rate is given by an
energy multiplied by a large scale Reynolds number. Note that a relation of this
type could only be derived for the total energy, not kinetic and magnetic
energies separately.

It is always true that $\epsilon_v + \epsilon_B \ge \epsilon_B$, and at large
$\mathrm{Rm}$, one expects the dissipation rate to be dominated by ohmic
dissipation, so that $\epsilon_v + \epsilon_B \approx \epsilon_B$, and one can
transform with little loss eq. (\ref{eq_Kolmogo_Etot}) into 
$\epsilon_B \le 2 e E_{\mathrm{tot}} \mathrm{Rm}$.
For infinite $\mathrm{Pm}$ (more precisely for 
$\mathrm{Rm}^2/\mathrm{Pm} \rightarrow 0$), one has the additional
simplification
\begin{equation}
\epsilon_B \le 2 e E_B \mathrm{Rm}
\label{eq_epsilon_EB}
\end{equation}
which suggests that in this case, the Kolmogorov phenomenology applies to the
magnetic field alone.

The next section will study among others the ratio $\epsilon_B/E_B$. For
infinite $\mathrm{Pm}$, $\epsilon_B/E_B$ is evidently bounded from above by $2 e
\mathrm{Rm}$. The bound is more complicated in the general case. Inserting
$\epsilon_v \ge \mathrm{Rm}^2/l_v^2$ into eq. (\ref{eq_epsilon_Etot}) leads to
\begin{equation}
\frac{\epsilon_B}{E_B} \le
2 e \mathrm{Rm} +2 e \frac{E_{\mathrm{kin}}}{E_B \mathrm{Pm}} \mathrm{Rm} +
(\| \nabla^2 \bm \Phi \| - \frac{1}{l_v^2}) \frac{\mathrm{Rm}^2}{E_B}
\label{eq_dissip_length}
\end{equation}
This expression frequently simplifies, as for instance for the G.O. Roberts flow,
because $\| \nabla^2 \bm \Phi \| = 1/l_v^2$ and the last term disappears. 

Up to
here, this section was kept in general terms independent of a particular choice
for $\bm \Phi$. The next section presents numerical simulations of the G.O.
Roberts dynamo, which means that $\bm \Phi$ is given by eq. (\ref{eq_flowI}),
which implies 
\begin{equation}
\| \nabla^2 \bm \Phi \| = 8 \pi^2
~~~,~~~
e=2\pi.
\end{equation}
The velocity is allowed to have a component independent of $z$ but is required
to be periodic with periodicity length 1 in the $x-$ and $y-$directions, so that 
$1/l_v^2 = 8 \pi^2$. The magnetic field on the other hand is allowed to have a
mean field along the $x-$ and $y-$directions, but is required to be strictly
periodic in the $z-$direction with periodicity length $l_z$, so that
$1/l_B^2 = (2 \pi/l_z)^2$.

\begin{figure}
\includegraphics[width=8cm]{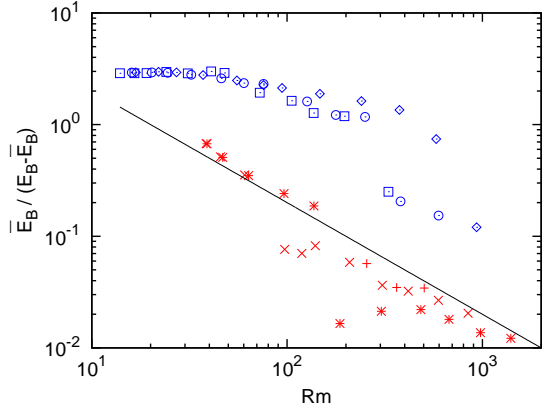}
\caption{(Color online)
${\overline E_B}/(E_B-{\overline E_B})$, as a function of $\mathrm{Rm}$ for
$l_z=2$ (blue symbols with an enclosed surface) and $l_z=1$ (red symbols built of crosses and stars).
For $l_z=2$, $\mathrm{Pm}$ is 0.3 (squares), 1 (circles) and 3 (diamonds). Results for the same 
$\mathrm{Pm}$ are shown for $l_z=1$: $\mathrm{Pm}=0.3$ (+), 1 (x) and 3 (*). The solid line
indicates the power law $\mathrm{Rm}^{-1}$.}
\label{fig_EB_mean}
\end{figure}

\begin{figure}
\includegraphics[width=8cm]{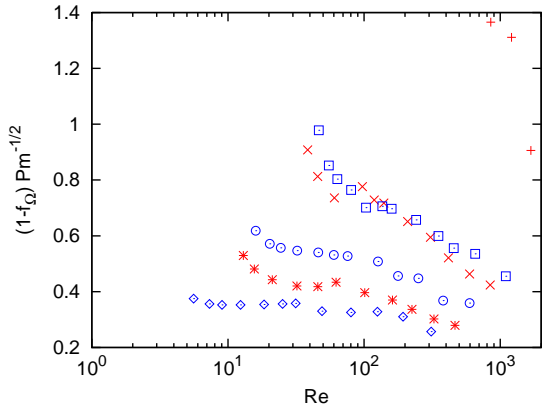}
\caption{(Color online)
$(1-f_\Omega) \mathrm{Pm}^{-1/2}$ as a function of $\mathrm{Re}$ with the same symbols
as in fig. \ref{fig_EB_mean}.}
\label{fig_f_Ohm}
\end{figure}

\section{Results}

We start with an overview of the solutions obtained numerically. The G.O. Roberts dynamo is
the prototypical $\alpha^2-$dynamo with a large mean field for control
parameters near a dynamo onset at small $\mathrm{Rm}$. We therefore compute both
the total magnetic energy density, $E_B$, and the energy density of the mean
field,
\begin{equation}
{\overline E_B} = \langle \frac{1}{V} A \int dz \frac{1}{2}  
\left(\frac{1}{A} \int dy \int dx \bm B \right)^2 \rangle
\end{equation}
where $A$ is the cross section of the computational volume $V$ in the
$x,y-$plane, $A=\int dx \int dy ~1$.
The ratio of mean to fluctuating energies, 
${\overline E_B}/(E_B-{\overline E_B})$,
is shown as a function of $\mathrm{Rm}$ in figure \ref{fig_EB_mean}. The
computations with the better separation of scales ($l_z=2$) show a magnetic
field dominated by the mean field at low $\mathrm{Rm}$. The contribution of the
mean field to the total field suddenly decreases at an $\mathrm{Rm}$ between a
few hundred and 1000. Beyond that sudden drop, ${\overline E_B}/(E_B-{\overline
E_B})$ decreases approximately as $1/\mathrm{Rm}$, whereas the runs with the
worse separation of scales ($l_z=1$) always have 
${\overline E_B}/(E_B-{\overline E_B})<1$ and show a decrease of this ratio
roughly in $1/\mathrm{Rm}$. A similar abrupt drop as in the case $l_z=2$ was
observed in convection driven dynamos \cite{Tilgne14b} where this drop signaled
a transition from dynamos generating a mean magnetic field to dynamos which do
not generate a mean field and in which all observed mean fields are merely
statistical fluctuations. In the convection dynamos, the transition occurred at
even smaller $\mathrm{Rm}$ than observed here for the G.O. Roberts dynamo. A
dramatic decrease of mean field generation at large $\mathrm{Rm}$ was already
noticed in \cite{Ponty11} and could be linked to a transition from a large
scale to a small scale dynamo. While the disappearence of the mean field is 
observed in the dynamos with columnar flow structures mentioned above, it 
does not appear in less organized turbulent flows \cite{Candel13}.
For the purpose of the present paper, it suffices
to note that the simulations discussed here contain cases of fields dominated
by a mean field as well as cases with a small mean field.

Another result relevant to the discussion below is the ratio of magnetic and
total dissipation rate,
\begin{equation}
f_\Omega = \frac{\epsilon_B}{\epsilon_B+\epsilon_v}.
\end{equation}
This ratio has received much attention in recent years, partly due to an attempt
to theoretically predict saturation field strengths in convection dynamos
\cite{Christ06}. $f_\Omega$ is zero for $\mathrm{Rm}$ below the onset
of dynamo action and tends to 1 for $\mathrm{Rm}$ tending to infinity.
Brandenburg \cite{Brande09} found that 
$\epsilon_v/(\epsilon_B+\epsilon_v) = 1 -f_\Omega$
scales as $\mathrm{Pm}^{1/2}$, and later \cite{Brande11} that
$\epsilon_v/\epsilon_B \propto \mathrm{Pm}^{0.6}$. Finally, a dependence
of the exponent on helicity was discovered in ref. \cite{Brande14}.
Figure \ref{fig_f_Ohm} plots $(1-f_\Omega) \mathrm{Pm}^{-1/2}$. The factor
$\mathrm{Pm}^{-1/2}$ obviously has not removed the $\mathrm{Pm}-$dependence from
the graph. In fact, there is no recognizable scaling for $f_\Omega$ in the
present data collection. The point of interest for the discussion below is that
$0.07 < f_\Omega < 0.8$ for all points in fig. \ref{fig_f_Ohm}.

\begin{figure}
\includegraphics[width=8cm]{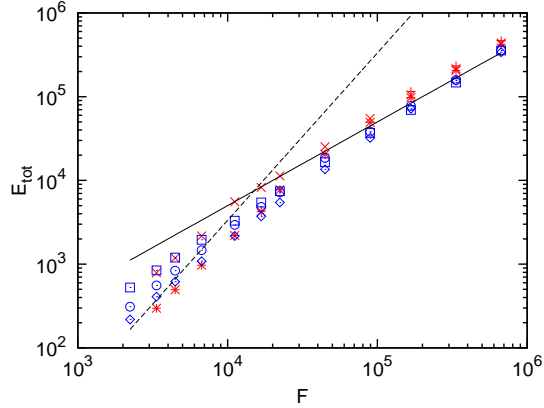}
\caption{(Color online)
Total energy as a function of the driving force $F$ with the same symbols as in fig. 
\ref{fig_EB_mean}. The continuous line shows the function $F/2$, the dashed line
is given by $F^2/(3\times 10^4)$.}
\label{fig_E_F}
\end{figure}

The most natural global quantity to investigate regarding its scaling behavior
is the total energy $E_{\mathrm{tot}}$ as function of the driving force $F$, as
shown in fig. \ref{fig_E_F}. The bounding theory of the previous section
provides us with an upper and a lower bound on $E_{\mathrm{tot}}$. According to
(\ref{eq_F_Etot}) and for $\mathrm{Rm} \rightarrow \infty$, one has 
for the G.O. Roberts flow $E_{\mathrm{tot}} \ge F / (4 \pi)$, 
and eq. (\ref{eq_F2_Etot}) yields 
$E_{\mathrm{tot}} \le F^2 \left(\frac{1}{\mathrm{Pm}}+\frac{1}{2}l_z^2 \right) / (128 \pi^4)$.
As mentioned above, the second bound is required by laminar solutions and can
only be of interest for small $F$. The
scaling for large $F$ suggested by the first bound
can also be inferred from heuristic arguments: The energy
injected into, and dissipated by, the flow is in order of magnitude $F U$ with
$U$ a characteristic flow velocity, whereas the total energy dissipation is
within the Kolmogorov phenomenology proportional to
$E_{\mathrm{tot}} U/L$, so that $E_{\mathrm{tot}} \propto F L$ with $L$ an
integral length scale and the proportionality factor is of course left
undetermined. This scaling matches the lower bound in eq. (\ref{eq_F_Etot}). The bounding
theory provides us with prefactors. In the high $\mathrm{Rm}$ limit, this
prefactor differs by a factor $2\pi$ from the prefactor obtained from the fit shown
in fig. \ref{fig_E_F}. The exponent on the other hand is close to the one found
in the actual simulations, in as far as one accepts exponents deduced from the
limited set of data in fig. \ref{fig_E_F}. The upper and lower bounds represent
the actual scalings in separate intervals of $F$, because one is based on 
laminar solutions, the other is not.

\begin{figure}
\includegraphics[width=8cm]{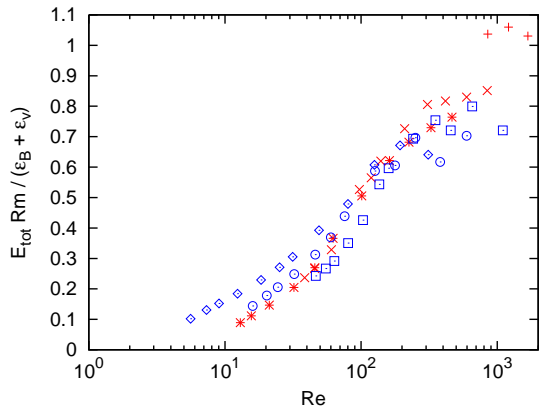}
\caption{(Color online)
$E_{\mathrm{tot}} \mathrm{Rm} /(\epsilon_B + \epsilon_v)$ as a function of
$\mathrm{Re}$ with the same symbols as in fig. \ref{fig_EB_mean}.}
\label{fig_E_dissip}
\end{figure}

\begin{figure}
\includegraphics[width=8cm]{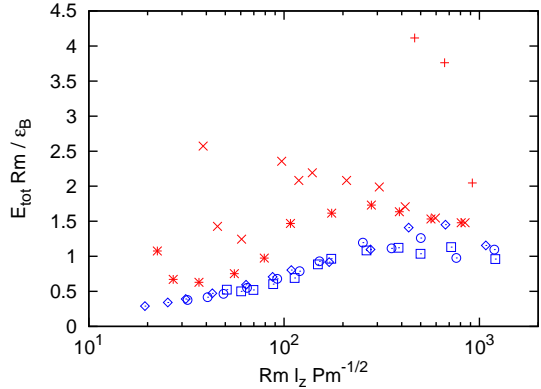}
\caption{(Color online)
$E_{\mathrm{tot}} \mathrm{Rm} / \epsilon_B$ as a function of
$\mathrm{Rm} l_z \mathrm{Pm}^{-1/2}$ with the same symbols as in fig. \ref{fig_EB_mean}.
The factor $l_z \mathrm{Pm}^{-1/2}$ is introduced on the $x-$axis for a better collapse 
of the data points and a more compact graphical representation, but it has no known 
physical significance.}
\label{fig_E_B_dissip}
\end{figure}

\begin{figure}
\includegraphics[width=8cm]{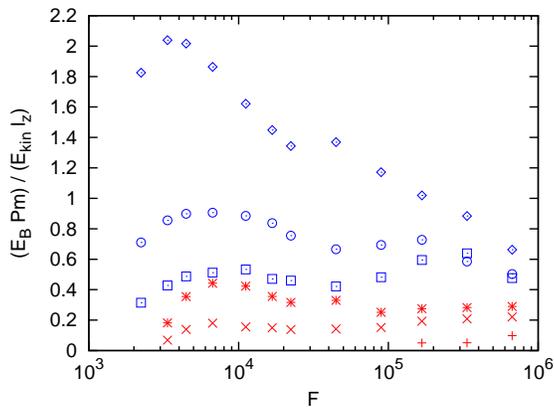}
\caption{(Color online)
$E_B \mathrm{Pm}/(E_{\mathrm{kin}} l_z)$ as a function of $F$ 
with the same symbols as in fig. \ref{fig_EB_mean}.}
\label{fig_equipartition}
\end{figure}

The upper bounds for the energy dissipation are not infested by laminar unstable
solutions. For large $\mathrm{Rm}$ (so that the term $8\pi^2 \mathrm{Rm}^2$ is
negligible in eq. (\ref{eq_epsilon_Etot})) one has
\begin{equation}
\epsilon_v + \epsilon_B \le 4 \pi E_{\mathrm{tot}} \mathrm{Rm}.
\end{equation}
The factor $4\pi$ is again an order of magnitude larger than the factor found in
the simulations at the highest $\mathrm{Rm}$, as seen in fig.
\ref{fig_E_dissip}. However, the important point about this bound is, in analogy
with the non-magnetic case \cite{Doerin02}, that it
provides a rigorous underpinning to the Kolmogorov phenomenology as long as it
is applied to total energy and dissipation in the form
$\epsilon_v + \epsilon_B \propto E_{\mathrm{tot}} U/L$.
The bounding theory does not in general provide support for any extensions of this
phenomenology in which magnetic and kinetic energies are split and which provides
us with a scaling for the magnetic field amplitude. 
Some simplification is achieved at large $\mathrm{Rm}$ in which case
$\epsilon_B \gg \epsilon_v$ and eq. (\ref{eq_epsilon_Etot}) simplifies to
\begin{equation}
\epsilon_B \le 4 \pi E_{\mathrm{tot}} \mathrm{Rm}.
\label{eq_E_B_dissip}
\end{equation}
Figure \ref{fig_E_B_dissip} verifies relation (\ref{eq_E_B_dissip}). A
regime asymptotic in $\mathrm{Rm}$ of the form 
$E_{\mathrm{tot}} \mathrm{Rm} / \epsilon_B \rightarrow \mathrm{const}$ cannot be
discernible yet because $\epsilon_B$
is not much larger than $\epsilon_v$ in any of these simulations.
Note the appearance of $E_{\mathrm{tot}}$ instead of 
$E_{\mathrm{kin}}/\mathrm{Pm}$ on the right hand side of eq. 
(\ref{eq_E_B_dissip}). An equation containing only magnetic quantities, at least 
in the form of a bound, is only obtained in the limit $\mathrm{Pm} \rightarrow
\infty$ (see eq. (\ref{eq_epsilon_EB})). Eq. (\ref{eq_def_epsilon}) always bounds $E_B$
in terms of $\epsilon_B$ but reflects the scaling pertaining to laminar
solutions, so that this bound will be of little practical interest in the
turbulent regime.

The distinction between $E_{\mathrm{tot}}$ and $E_{\mathrm{kin}}/\mathrm{Pm}$ on
the right hand side of eq. (\ref{eq_E_B_dissip})
becomes mute if there is equipartition between magnetic and
kinetic energies, because then, all energies are proportional to each other. We
are therefore led to investigate the ratio $E_B \mathrm{Pm}/E_{\mathrm{kin}}$ in
fig. \ref{fig_equipartition}. This ratio is not perfectly constant as $F$ is
varied, but except
for the case $l_z=2$, $\mathrm{Pm}=3$, it varies around its mean by less than 30\%
(and less than a factor of 2 over all). In the range of $\mathrm{Rm}$ in which
fig. \ref{fig_E_B_dissip} suggests the validity of a Kolmogorov phenomenology
(typically the last three or four points of each series), the ratio of energies
varies by at most 20\% around its mean value, while the energies themselves vary
by more than an order of magnitude. For the purpose of the scaling implied by eq.
(\ref{eq_E_B_dissip}), $\epsilon_B \propto E_{\mathrm{tot}} \mathrm{Rm}$, we
have nearly $E_B \propto E_{\mathrm{kin}}/\mathrm{Pm} \propto E_{\mathrm{tot}}$
and it cannot be tested whether $E_{\mathrm{tot}}$ on the right hand side of this
scaling relation is essential or whether it could be replaced by $E_B$.

However, it is clear from fig. \ref{fig_equipartition}
that $E_B \mathrm{Pm}/E_{\mathrm{kin}}$ depends on $l_z$
and $\mathrm{Pm}$, so that there is no strict equipartition between magnetic and
kinetic energies, but only a proportionality between the two. The same situation
was already found in convection dynamos, in which $E_B \propto E_{\mathrm{kin}}$
at high $\mathrm{Rm}$, with a proportionality factor depending on additional
control parameters \cite{Tilgne12b}. Ref. \cite{Ponty11} claims to have
found equipartition between $E_B$ and $E_{\mathrm{kin}}$. However, this finding
was fortuitous because it was obtained for $l_z=2$ and $\mathrm{Pm}$ of order 1.
For these parameters, one indeed obtains $E_B \approx
E_{\mathrm{kin}}/\mathrm{Pm}$, but not at general $l_z$ and $\mathrm{Pm}$. From
a balance between the Lorentz force term $(\nabla \times \bm B) \times \bm B$
and the advection term $\bm v \cdot \nabla \bm v/\mathrm{Pm}$ and the assumption
that $\bm B$ and $\nabla \times \bm B$ are
dominated by a mean field which varies on the length scale $l_z$ while $\bm v$
varies on the length scale 1, one obtains
$E_B \propto l_z E_{\mathrm{kin}}/\mathrm{Pm}$. The same result was derived by
Brandenburg \cite{Brande01} from an argument based on helicities and the
assumption of a dynamo field dominated by its mean component. While these
arguments do yield a non-trivial prefactor in $E_B \propto 
E_{\mathrm{kin}}/\mathrm{Pm}$, they require a strong mean field which does not
exist in all simulations presented here (see fig. \ref{fig_EB_mean}),
so that it comes to no surprise that
$E_B \mathrm{Pm} / (E_{\mathrm{kin}} l_z)$ does not yield a value independent of
all other parameters in fig. \ref{fig_equipartition}.

Let us neglect these additional dependences for the moment and assume 
$E_{\mathrm{kin}}/\mathrm{Pm}$ to be proportional to $E_B/l_z$. Since
$E_{\mathrm{kin}}/\mathrm{Pm}$ is for most points in fig.
\ref{fig_equipartition} larger than $E_B$, let us approximate
$E_B + E_{\mathrm{kin}}/\mathrm{Pm} \approx b E_B/l_z$
with some numerical constant $b$, so that eq. (\ref{eq_Kolmogo_Etot}) becomes
\begin{equation}
\epsilon_v + \epsilon_B \le b 2 e \frac{E_B}{l_z} \mathrm{Rm}.
\label{eq_ineq_b}
\end{equation}
For large $\mathrm{Rm}$, one expects $\epsilon_B \gg \epsilon_v$. The inequality
\begin{equation}
\epsilon_B \le b 2 e \frac{E_B}{l_z} \mathrm{Rm}
\label{eq_ineq_b2}
\end{equation}
derived from relation (\ref{eq_ineq_b}) should therefore not be much more
inaccurate than relation (\ref{eq_ineq_b}) itself.

\begin{figure}
\includegraphics[width=8cm]{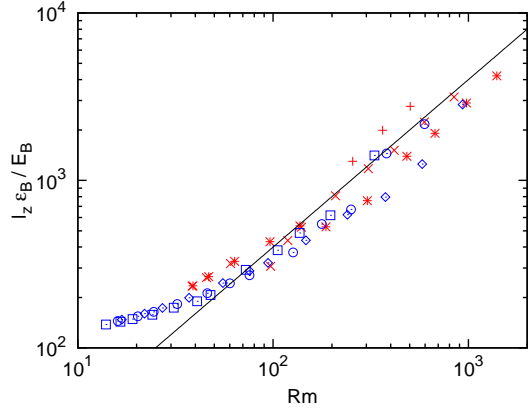}
\caption{(Color online)
$l_z \epsilon_B/E_B$ as a function of $\mathrm{Rm}$ with the same symbols as in
fig. \ref{fig_EB_mean}. The solid line shows the function $4~ \mathrm{Rm}$.}
\label{fig_lB}
\end{figure}

Fig. \ref{fig_lB} tests eq. (\ref{eq_ineq_b2}). At the level of accuracy of double logarithmic plots, the scaling 
$\epsilon_B \propto \frac{E_B}{l_z} \mathrm{Rm}$ implied by (\ref{eq_ineq_b2})
viewed as an equality adequately represents the data cloud at large
$\mathrm{Rm}$. It should be stressed that this scaling is not strictly supported
by the bounds of the previous section, since some heuristic input was necessary
concerning $E_B \mathrm{Pm} / E_{\mathrm{kin}}$.

It is of interest to characterize the magnetic field through a time or length
scale extracted from its energy and dissipation rate, $\epsilon_B/E_B$. An
initial study on convection dynamos found $\epsilon_B/E_B \propto \mathrm{Rm}$
\cite{Christ04}. Later investigations yielded smaller exponents \cite{Tilgne14b}.
The present computations do not reach to high enough $\mathrm{Rm}$ to accurately
determine an exponent, but fig. \ref{fig_lB} suggests 
$\epsilon_B/E_B \propto \mathrm{Rm}$. Eq. (\ref{eq_dissip_length}) applied to
the G.O. Roberts flow reads
\begin{equation}
\frac{\epsilon_B}{E_B} \le 4 \pi \mathrm{Rm} \left( 1+
\frac{E_{\mathrm{kin}}}{E_B \mathrm{Pm}} \right).
\end{equation}
If the bracket tends to a constant for $\mathrm{Rm} \rightarrow \infty$, this
inequality shows that the exponent $c$ in 
$\epsilon_B / E_B \propto \mathrm{Rm}^c$ must obey $c \le 1$.

\section{Conclusion}

The G.O. Roberts dynamo is known as a dynamo generating a mean magnetic field,
but this study has shown that as $\mathrm{Rm}$ is increased, there is a sudden
drop in the contribution by the mean field to the total magnetic energy. The
same phenomenon was already observed in convection driven dynamos in plane
layers, which perhaps is not surprising because both flows consist of helical
vortices with parallel axes. However, if this is the essential feature common to
both flows, one would expect an analogous behavior in convection driven dynamos
in spherical shells. If the analogy is valid, the energy of the axisymmetric
modes should suddenly decrease at high $\mathrm{Rm}$ in favor of modes with a
non zero azimuthal wavenumber.

It is possible to derive several rigorous bounds for the G.O. Roberts dynamo.
These concern the total energy, either as a function of the driving force or the
total dissipation. The same functional dependences appear in these bounds as in
heuristic arguments, but the prefactors in proportionalities are determined in
the bounds and differ by about an order of magnitude from prefactors determined
from best fits to the numerical data. The prefactors and numerical constants in
the bounds presented above could be improved with some numerical effort. 
The Kolmogorov phenomenology relating
total energy to total dissipation is compatible with the bounds. But the
optimum theory as presented here does not give any bound of interest for the
turbulent regime on magnetic energy
alone, unless one is interested in the limit of infinite
$\mathrm{Pm}$, or unless one accepts an independent result, extracted from
numerical simulations, which states that the ratio of magnetic and kinetic
energy is nearly independent of $\mathrm{Rm}$ at large $\mathrm{Rm}$. In these
cases, one obtains a phenomenology of the Kolmogorov type for the magnetic field
alone.

\acknowledgments
The author acknowledges support from the Helmholtz-alliance ``Liquid Metal Technologies''.


%

\end{document}